# Endwall and leading-edge film cooling of turbine blades in a hydrogen-fueled rotating detonation combustor-turbine coupled system

Yeqi Zhou[1], Songbai Yao[1,2,*], Jingtian Yu[1,2], Weijia Qian[3], Ping Wang[3], Wenwu Zhang[1,2]

[1]Ningbo Institute of Materials Technology and Engineering, Chinese Academy of Sciences, Ningbo 315201, China

[2]University of Chinese Academy of Sciences, Beijing 100049, China

[3]Institute for Energy Research, Jiangsu University, Zhenjiang 212013, China

**Abstract**

This study performs a three-dimensional numerical simulation of the coupled flow field in a hydrogen-air rotating detonation combustor (RDC)-turbine system to evaluate the effectiveness of different film cooling strategies for the turbine blades. The results demonstrate that combining the endwall cooling with leading-edge film cooling effectively reduces blade surface temperatures while improving turbine flow field stability and blade protection. For endwall cooling, numerical simulations compare circular and slot hole configurations. Circular holes consume less cooling air than slot holes while maintaining comparable cooling performance, making them the preferred choice. For the leading-edge film cooling, both the vertical and the vertical-inclined schemes are examined. The vertical-inclined scheme demonstrates higher cooling efficiency and improved secondary flow attachment, ensuring greater stability under the oscillatory effects of the detonation flow. Additionally, the flow fields of film-cooled turbine blades with and without the propagation of the rotating detonation wave are compared, revealing that the upstream rotating detonation flow field facilitates the downstream diffusion of secondary film cooling jets.

**Keywords**: Rotating detonation combustor; Hydrogen; Turbine blade; Film cooling; Endwall cooling; Leading-edge cooling

## 1. Introduction

The rotating detonation engine (RDE) is a revolutionary propulsion technology that generates thrust through continuous circumferential propagation of detonation waves, which is distinguished by its pressure gain effect, higher thermodynamic efficiency, and rapid heat release characteristics [1-3]. Unlike conventional engines based on deflagration combustion, RDEs utilize annular or cylindrical combustors where rotating detonation waves (RDWs) instantaneously combust fuel-oxidizer mixtures, significantly reducing engine complexity, lowering weight and improving fuel efficiency [4-6]. This technology has been extended to applications such as rotating detonation rocket engines, air-breathing propulsion systems, and gas turbine integration [7-19].

Among these, turbine-based rotating detonation combustors (RDCs) have undergone continuous development. The compact design and rapid energy conversion of RDEs provide a promising approach for developing efficient, power-dense gas turbine engines, particularly in aerospace applications where space and weight are constrained. For example, Wolanski et al. [20] evaluated the stability of RDC operation in three combustion chamber configurations: open-flow, throttled, and GTD-350 gas turbine-integrated setups. Their findings indicated that when the GTD-350 engine operated with gaseous hydrogen or Jet-A/hydrogen dual-fuel modes, thermal efficiency improved by 5–7% compared to the baseline engine. Huff et al. [21, 22] proposed a radial RDC integrated with a centrifugal turbine. Experimental results showed power outputs exceeding 70 kW at rotational speeds above 110 kRPM, with combustion chamber and turbine guide vane interactions causing approximately 40% total pressure loss. The peak system thermal efficiency reached 40%. Naples et al. [23] integrated an RDC into a T63 gas turbine and compared its performance to conventional combustors. The RDC exit exhibited pressure fluctuations several times greater than those observed in conventional systems. Turbine isentropic efficiency decreased by 1.3%, while the circumferential temperature nonuniformity coefficient remained comparable to conventional combustors. Fievisohn et al. [24] successfully demonstrated closed-loop RDC operation in a T63 gas turbine, with an output power reaching 82% of the rated capacity. Su et al. [25] investigated the flow characteristics and performance of supersonic turbine stages in RDCs under varying rotational speeds, stator-to-rotor ratios, and hub radii. Their study classified stator operational modes into non-wave contact modes, opposite-side λ-wave oblique shock modes, and derived opposite-side λ-wave oblique shock modes. Zhao et al. [26] conducted two-dimensional numerical simulations to analyze RDC-turbine interactions and proposed a control device incorporating a deflecting wedge to automatically regulate detonation wave propagation. Zhou et al. [8] experimentally compared the performance of axial and radial turbines integrated with RDCs, demonstrating that the axial configuration outperformed the radial design. Zhou et al. [27] contributed by experimentally assessing the performance of axial and radial turbines integrated with RDCs, demonstrating that the axial turbine configuration outperforms its radial counterpart. Zhang et al. [28] performed numerical studies on the interaction between continuously rotating detonation waves and turbine stator blades, elucidating the dynamic processes when detonation waves propagate in multiple directions. Shao et al. [29] employed two-dimensional numerical simulations with methane recuperation gas to investigate the coupling between the RDC and turbine stator blades, revealing that increased methane conversion induced a transition from single-wave to unstable detonation modes. Rathod and Meadows [30] integrated an RDC with industrial NGVs using a nonoptimized straight duct design and employed 3D numerical simulations to assess how shock-induced unsteady flows and nonuniform temperature profiles contributed to total pressure losses and irreversibility. Grasa and Paniagua [31] investigated the design and optimization of diffusive stator vanes to achieve reductions in pressure loss and rotor forcing while preserving a broad operating envelope. Qi et al. [32] established a cycle calculation model for a 25 MW RDC turbine to evaluate cycle performance and component characteristics under both design and off-design conditions.

*Corresponding author: yaosongbai@nimte.ac.cn (S. Yao)

The rapid energy release associated with detonation combustion, coupled with high-frequency shock wave impingement, exposes engine components—such as combustor walls, injectors, and nozzles—to peak temperatures that can exceed material melting points and cause thermal fatigue. For example, Bykovskii and Vedernikov [33] identified that the peak heat flux of the RDC could exceed 9 MW/m², and Micka et al. [34] reported heat fluxes up to 25 MW/m². Similar heat flux levels have also been reported in previous studies [35, 36]. Therefore, it becomes one of the most challenging issues for the engineering application of the RDC. Previous studies have implemented water-cooling channels for the RDC [5, 37, 38]. Among these, Glaubitz et al. [37] achieved 60 seconds of sustained rotating detonation, whereas Teasley et al. [5] implemented a combined thermal protection system integrating water-cooling and regenerative cooling (using liquid methane) in the RDC. The engine demonstrated cumulative operational durations exceeding ten minutes. Kawalec et al. [39] employed liquid $N_2O$ in a sounding rocket RDC, allowing stable operation for up to 4 s.

Due to the rapid heat release rate of the RDC, a high-temperature, high-pressure flow field forms downstream of the combustor, which imposes demanding cooling requirements on the turbine downstream. Film cooling technology is an effective and widely used cooling method for turbine blades in aero-engines [40-43]. Tian et al. [44] identified RDW blockage and oblique shock wave-modulated oscillations, with cooling flow reducing peak RDW pressure by 6–8% without affecting propagation velocity. Yu et al. [45] evaluated the cooling efficiency of the film holes under varying mass flow rates of the primary and secondary flows. They also found that the use of cat-ear-shaped cooling holes could improve efficiency of the film-cooled RDC [46, 47]. Li et al. [48] integrated film cooling with nozzle design in an RDC, and Shen et al. [49] analyzed the influence of the angle of cylindrical holes. Liu et al. [50] analyzed turbine blade film cooling under the inflow conditions of a two-dimensional RDC using numerical simulations. They found that the detonation inflow caused uneven pressure distribution and increased aerothermal loads, which disrupted conventional film cooling and reduced its effectiveness. Sridhara et al. [51] investigated a film-cooled RDC with 480 cylindrical film cooling holes and found that using air as a coolant resulted in increased heat release, which was attributed to secondary combustion.

Although considerable research has been conducted on the fundamental characteristics of film cooling on RDC walls, the critical issue of thermal protection for downstream turbine blades integrated with an RDC remains significantly underexplored. Prior studies, many of which rely on simplified two-dimensional models, have highlighted the intricate flow dynamics and severe thermal loads experienced within the turbine region when coupled with an RDC. This gap in understanding poses challenges for the practical implementation and reliability of such systems. Therefore, this study aims to address this crucial limitation by investigating various film cooling strategies—alongside a novel combined cooling scheme—specifically designed to enhance the thermal protection of turbine blades in a fully three-dimensional RDC-turbine system fueled by hydrogen and air. The focus is placed on effectively protecting both the endwalls and leading edges of the blades, which are particularly vulnerable to high thermal stresses, thereby advancing the potential for improved durability and performance of turbine components in RDC-integrated engines.

## 2. Numerical modeling
### 2.1 RDC model integrated with turbine blades

The computational model in this study divides the domain into two regions: the rotating detonation region (combustor) upstream and the turbine (stator blades) region downstream. In the turbine region, inner and outer endwall film cooling and leading-edge film cooling are implemented on the endwall surfaces and stator blades, respectively. The parameters and operating conditions of the turbine blades are listed in Table 1. Figure 1(a) illustrates the RDC flow field used in the numerical simulation, with the inner and outer wall diameters set to 30 mm and 40 mm, respectively, and the combustor length set to 40 mm. Because of the circular symmetry of the combustor domain, a structured mesh is implemented. Figure 1(b) displays the turbine stator blade model employed in this study. To enable seamless transfer of the combustor flow field to the turbine blades, the hub and blade tip diameters are also configured to 30 mm and 40 mm, with a blade height of 10 mm. An unstructured mesh is used for the turbine flow field. The number of circumferential blades is set to 12, and a periodic boundary condition is applied in ICEM CFD to arrange the single-blade mesh into a periodic array, resulting in a full-annular configuration. The grid interface connects the flow field exit of the RDC to the inlet of the turbine blades, generating the integrated computational domain shown in Figure 1(c).

Table 1. Turbine blade geometric parameters and operating conditions

| | | |
|---|---|---|
| Geometric parameters | Blade height | 5 mm |
| | Airfoil area | 8.22542 mm² |
| | Chord length | 7.32399 mm |
| | Meridional length | 6 mm |
| | Stagger angle | 35.0° |
| | Aspect ratio | 0.6827 |
| Operating Conditions | Reynolds number | $1.03\times10^5$ |
| | Mach number | 0.3 |
| | Angle of attack | 5° |

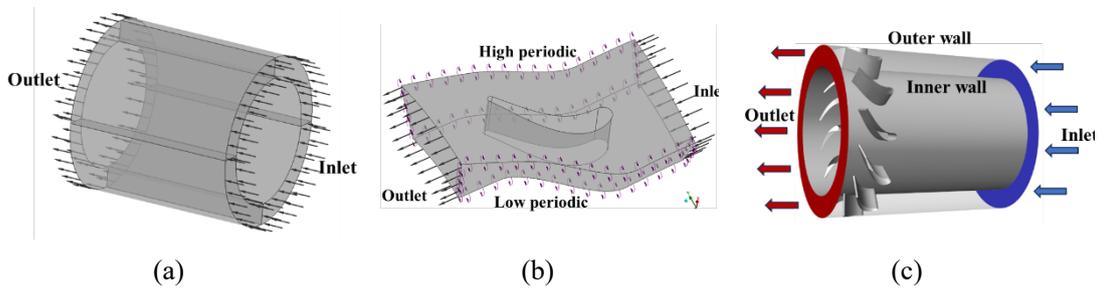

(a)          (b)          (c)

Figure 1. Schematic of the computational domain: (a) Detonation zone, (b) Turbine blade, and (c) Entire coupled region.

For the film cooling system on the endwalls, two distinct configurations are implemented: slot holes and circular holes. Figure 2(a) illustrates the blade model with film cooling slots on the endwalls. Film cooling slot holes are incorporated into the endwall surfaces near the hub and blade tip regions of each blade. The outer slot holes have a thickness of 0.4 mm and an axial length of 3.7 mm, while the inner slot holes have the same thickness (0.4 mm) with a reduced axial length of 3.1 mm. Secondary cooling flow is introduced into the slot passages through two annular collector chambers before merging with the mainstream flow. The slot hole position is designed to alleviate the higher heat load at the horseshoe vortex and is inclined at 30° to the wall surface to ensure the attachment of the secondary flow. Additionally, it is also inclined at a 30° angle with the tangent of the arcuate wall surface, ensuring that the direction of the secondary flow is as consistent as possible with the mainstream flow. Figure 2(b) shows the endwall film cooling system with discrete holes, in which circular holes with a diameter of 0.4 mm are positioned on the endwall surfaces. Eight circular holes are located on the outer side, and seven circular holes are located on the inner side. The angle and position of the holes are aligned with those of the slot holes. However, due to the smaller passage area of the holes, their axial length is slightly longer than that of the slot holes. To improve the computational accuracy of the endwall film cooling slot/circular hole regions, a local mesh refinement factor of 3 is applied to ensure a finer mesh distribution. Additionally, five layers of mesh are used at the blade tip and hub as boundary layers, with a transition ratio of 1.2. The thickness of the first boundary layer is set to 0.12 mm. The boundary layer refines the flow field at the junction of the secondary and mainstream flows.

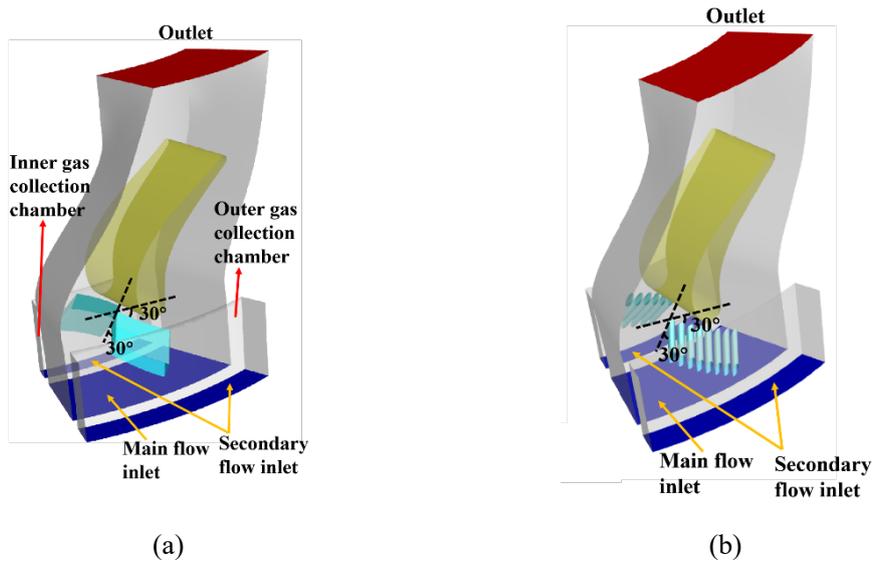

(a)                         (b)

Figure 2. Schematic of inner-outer endwall film cooling: (a) Slots holes and (b) Circular holes.

Two distinct design schemes for the leading-edge film cooling holes are examined. The first design uses traditional vertical holes, while the second employs combined holes to enhance the cooling effect along the middle section of the leading edge. Both configurations consist of three rows of film cooling holes at the blade's leading edge, as shown in Figure 3(a). The precise locations of the holes on the blade, along with the positions of the monitoring lines, are shown in Figure 3(b). The cooling system is connected to a cylindrical collector chamber, which supplies cooling air to the film cooling holes. The hole orientations for both designs are illustrated in Figures 3(c) and 3(d), respectively.

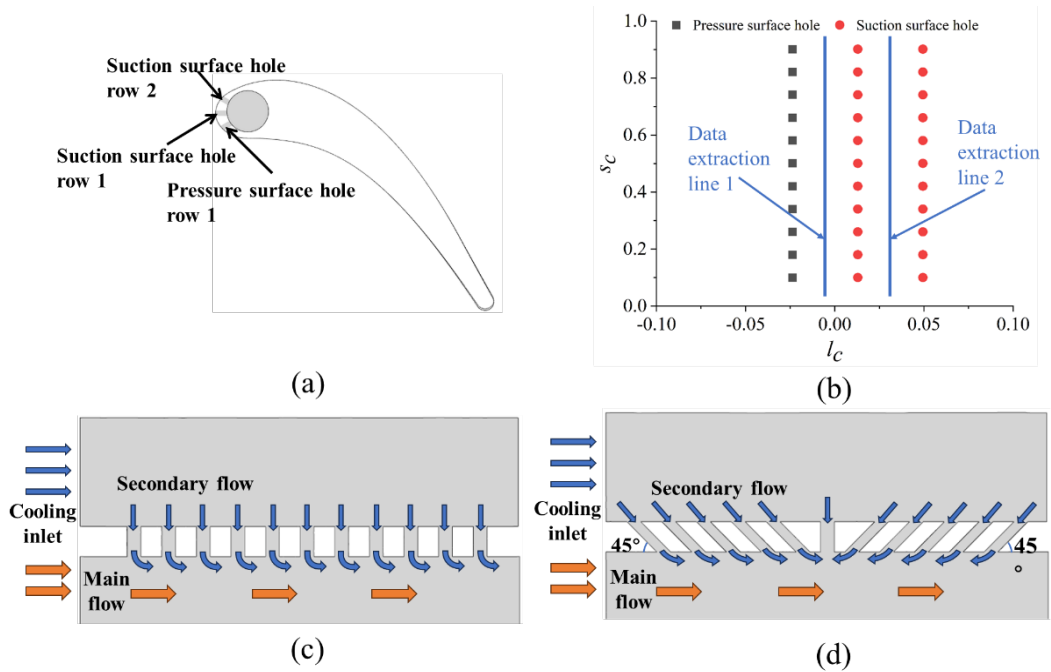

Figure 3. Schematic of leading-edge film cooling: (a) Arrangement of hole rows, (b) Film cooling hole layout, (c) Vertical-angle scheme, and (d) Vertical-inclined scheme.

**2.2 Boundary conditions**

The combustor inlet is specified as a mass flow inlet, with a stoichiometric hydrogen-air premixed gas entering at a mass flow rate of 100 g/s and a total temperature of 300 K, with a turbulence intensity of 5%[52]. The species transport model is employed, and a hydrogen-air one-step reaction mechanism is used to simulate the hydrogen-air premixed combustion process. Consistent with previous studies [53-55], we use a species transport model with a finite-rate kinetics. An ignition region is positioned at the combustor head, as shown in Figure 4a, with a pressure of 2 MPa, an initial velocity of 1800 m/s, and an initial temperature of 2000 K. A fuel pre-fill region is placed upstream of the ignition region to guide the detonation wave. The combustor wall is specified as an adiabatic no-slip wall, and the outlet is connected to the turbine inlet through a grid interface. The turbine outlet is defined as a pressure outlet with a pressure of 0.1 MPa. The endwall cooling gas collector chamber inlet is set as a pressure inlet of 0.65 MPa. The leading-edge cooling collector chamber inlet is defined as a mass flow inlet with a flow rate of 3 g/s and an initial temperature of 300 K for the cooling gas.

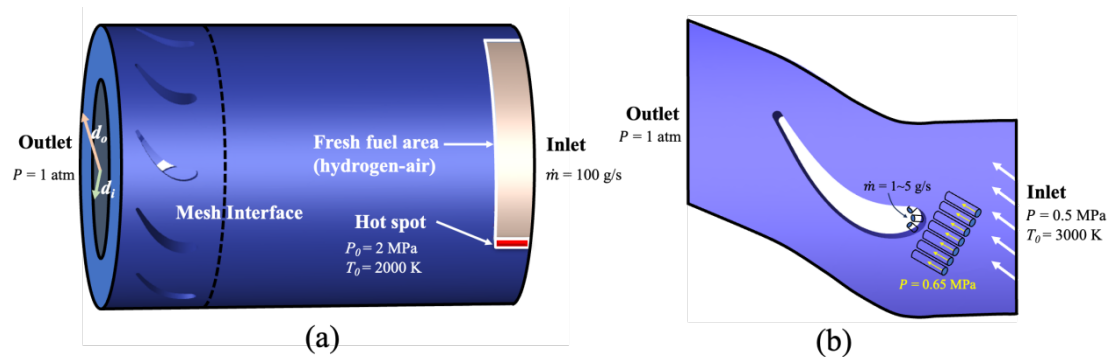

Figure 4. Configuration of initial and boundary conditions: (a) Fully coupled RDC and (b) Single-blade model.

For comparison, a simplified single-blade turbine flow field simulation is conducted as a benchmark case for the non-detonation flow field, as illustrated in Figure 4b. A pressure inlet of 0.5 MPa is employed to approximate the interface between the turbine region and the detonation combustor. The inlet direction is perpendicular to the axial direction, with a 30° angle relative to the normal direction, simulating the velocity direction at the exit of the detonation flow field. The mainstream temperature is set to 3000 K to replicate the extreme conditions of high-temperature regions induced by shock waves impinging on the blade. The endwall cooling gas collector chamber inlet is defined as a pressure inlet of 0.65 MPa, while the leading-edge cooling collector chamber inlet is specified as a mass flow inlet with a mass flow rate of 1–5 g/s and an initial temperature of 300 K for the cooling gas. Periodic boundary conditions are applied to the lateral boundaries of the flow field, and the remaining boundaries are treated as adiabatic, no-slip walls.

**2.3 Calculation methods and verification**

The density-based solver is used to solve the continuity, momentum, energy, and species transport equations within the framework of the two-dimensional, unsteady Navier-Stokes equations, implemented in ANSYS Fluent. The chemical reaction is modeled using a global one-step reaction, where the reaction rate is computed using the Arrhenius equation with model parameters provided in Ref. [56]. Spatial discretization is performed using a second-order upwind scheme. The convective flux is decomposed using the AUSM (Advection Upstream Splitting Method). For turbulence modeling, the realizable k–ϵ turbulence model is utilized. Compared to the standard and RNG k–ϵ models, the realizable k–ϵ model is better suited for separated flows and flows with complex secondary flow features. To handle the near-wall flow, non-equilibrium wall functions are applied, which are particularly effective in the presence of large pressure gradients.

For the grid sensitivity analysis, grid resolutions of 0.25 mm, 0.3 mm, and 0.35 mm are used to discretize the computational domain of the detonation region. The temperature contours of the stabilized RDW flow fields, obtained at different resolutions, are shown in Figure 5. A pressure monitoring point is located 8 mm from the inlet, and the pressure profiles for each case are also displayed. It is observed that both the flow fields and pressure profiles are nearly identical across the cases. The computed propagation speed of the RDW deviates by less than 7% from the theoretical Chapman–Jouguet value, as predicted by NASA's Chemical Equilibrium Applications (CEA) program [57]. Based on these results, the 0.3 mm grid is selected for the subsequent simulations of the detonation region. To validate the numerical setup, chemical kinetics, and turbulence model developed for predicting film cooling efficiency, Appendices A, B, and C present numerical calculations of benchmark cases compared with experimental data and theoretical analysis.

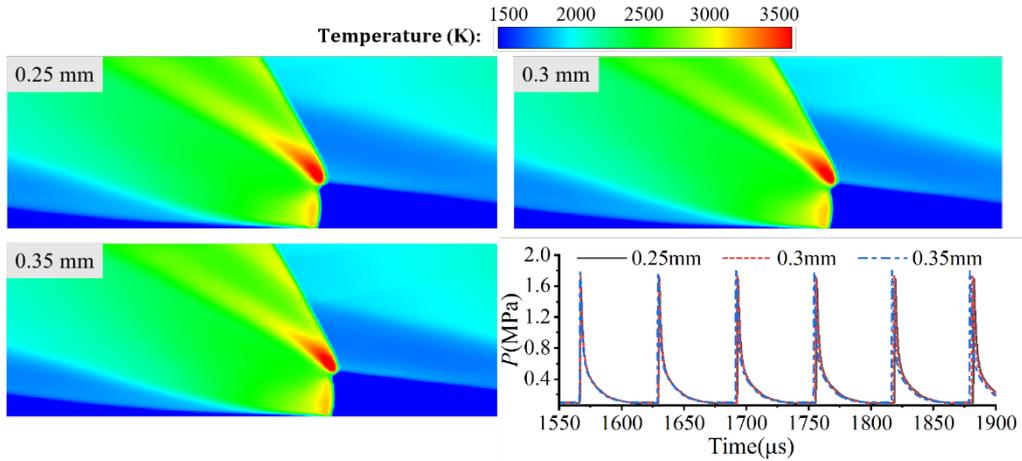

Figure 5. Flow field pressure contours and pressure profiles at different resolutions.

## 3. Results and discussion
### 3.1 RDC-turbine flow field without film cooling

Firstly, a numerical simulation of the coupled flow field without film cooling is performed to investigate the interaction between the shock wave flow field and the turbine stator blades. Figure 6 illustrates that the shock wave interaction with the turbine blades exhibits several key characteristics. After the oblique shock wave impinges on the turbine stator blade, a reflected wave propagates upstream, leading to a temperature increase in the affected region. Moreover, the shock wave amplitude decreases, and the oblique shock wave impacts the leading edge of the turbine stator blade, generating a high-temperature, high-pressure region. The downstream turbine flow field exhibits a relatively lower temperature. These key characteristics align with findings from previous studies on the RDC coupled with turbine blades [28, 58].

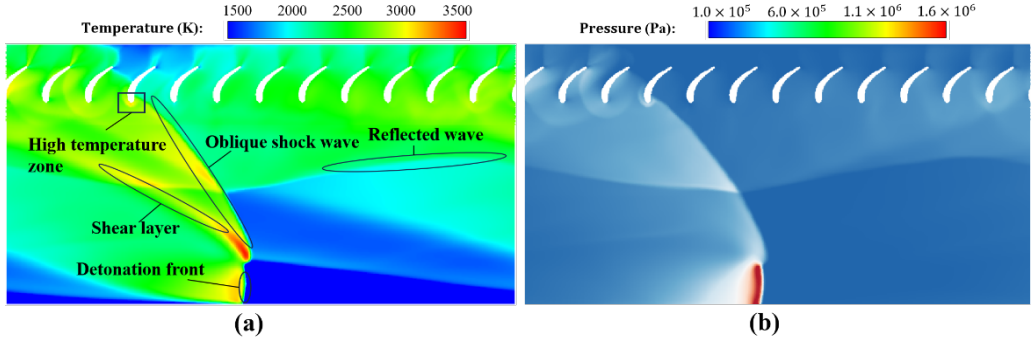

Figure 6. Contour plots of the coupled flow field: (a) Temperature and (b) Pressure.

Figure 7 depicts the flow field evolution after the oblique shock impinges on the turbine blades. The oblique shock first generates a reflected wave at the blade's leading edge. As the reflected wave propagates, it further reflects off the suction or pressure surfaces of adjacent blades, forming a λ-shaped wave. As the λ-shaped wave propagates downstream, secondary shock waves form due to reflections and interactions with other waves, leading to a more complex wave system. After the airflow passes through the turbine blades, separation occurs due to changes in pressure and velocity, creating a swallowtail wave at the blade's trailing edge. The reflected wave exerts intense impact on the turbine blades, leading to blade vibration and potentially increasing mechanical loads. Additionally, the vortices generated by the trailing flow effects behind the turbine blades further disturb the flow field. In summary, these complex shock waves concentrate thermal load on the turbine blades, thereby reducing aerodynamic stability.

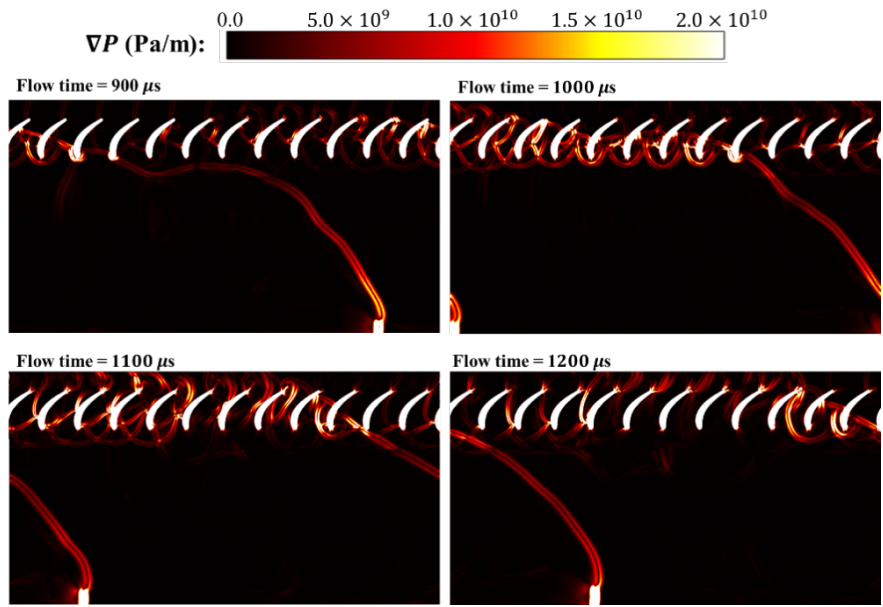

Figure 7. Pressure gradient contour plot of the interaction between the oblique shock wave and the turbine.

To evaluate the impact of the detonative flow field on the temperature distribution of the stationary blade, the region near the blade is divided into five areas: pressure surface, suction surface, leading-edge horseshoe vortex, trailing-edge wake vortex, and near-inviscid flow region. As shown in Figure 8(a), the highest thermal loads are concentrated in the horseshoe vortex region, followed by the pressure and suction surface walls and the trailing-edge wake vortex. Figure 8(b) illustrates the gas flow direction within the turbine. Upstream of the blade, the mainstream flow forms an angle of approximately 30° with the Z-axis, which increases to about 45° after passing through the blade. This flow field analysis provides a basis for optimizing cooling hole placement. For instance, wall cooling holes are positioned at the leading edge of the horseshoe vortex and inclined at 30° to the circumferential direction to align the secondary flow with the mainstream flow as closely as possible.

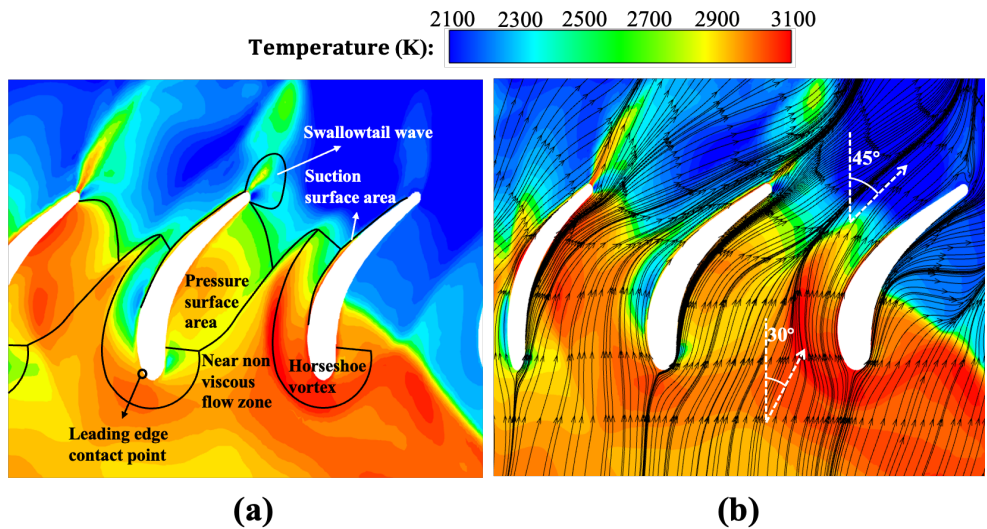

Figure 8. Flow field near the blade: (a) Local zone and (b) Streamlines.

**3.2 Flow field analysis with endwall film cooling on the turbine blades**

The turbine endwalls, positioned between the blades and the casing, are directly exposed to high-temperature airflow. Both the inner and outer endwalls withstand substantial thermal loads, particularly in high-temperature, high-pressure environments. Unlike turbine blades, endwalls lie along the periphery of the airflow, subjecting them to steeper temperature gradients and higher heat flux densities.

Figure 9(a) illustrates the blade temperature distribution under shock wave impingement, revealing significant thermal loading near the endwalls, where high-temperature gas accumulates. To mitigate this, an endwall cooling scheme is proposed for thermal protection. Based on the heat load distribution in Figure 8, cooling holes are strategically placed on the endwalls ahead of each blade, targeting the horseshoe vortex region to reduce localized thermal stress.

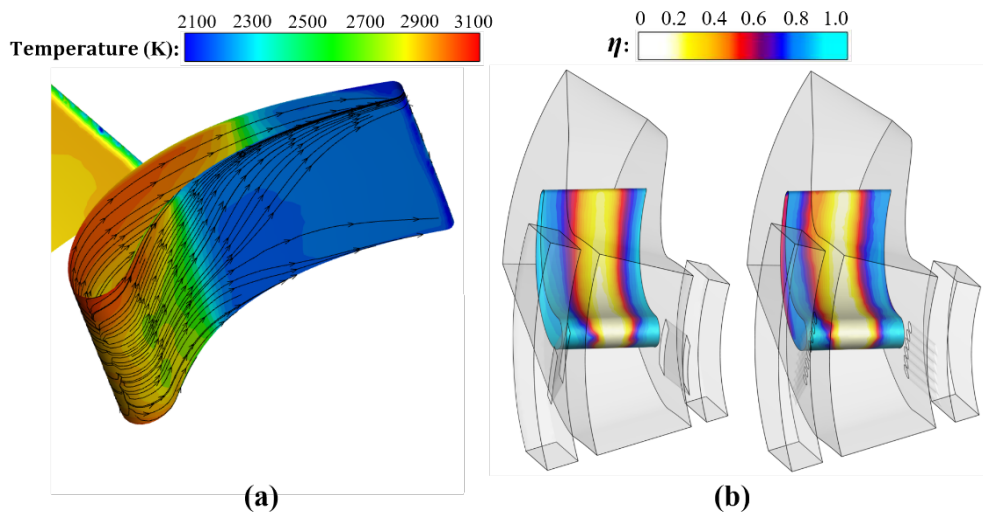

Figure 9. Temperature contour plot of (a) an uncooled blade and (b) a single blade with circular-hole- or slot-hole-based endwall film cooling.

We evaluate two endwall film cooling designs: slot and circular holes. To assess their feasibility, a single-blade model is used for analysis. The film cooling efficiency [41] is defined as

$$\eta = \frac{T_{\mathrm{main}} - T_{\mathrm{coolant}}}{T_{\mathrm{main}} - T_{\mathrm{aw}}} \quad (1)$$

where $T_{\mathrm{aw}}$ is the wall temperature, $T_{\mathrm{main}}$ is the mainstream temperature, and $T_{\mathrm{coolant}}$ is the coolant temperature at the inlet. Figure 9(b) presents the cooling effectiveness of both designs under the same steady inlet conditions. The inlet temperature and pressure are obtained from the peak values at the exit of the RDC. In the immediate region downstream of the endwall film hole exit, the film cooling efficiency is close to unity because the wall is almost entirely covered by undiluted coolant that has just emerged from the hole. At this location, the coolant temperature is still essentially the same as its supply temperature, with minimal heating from the hot mainstream flow. Consequently, according to the definition of $\eta$, the region immediately downstream of the film-cooling hole exit exhibits a value of $\eta$ close to 1. A similar phenomenon was reported in the experimental study [59, 60]. The slot hole design achieves an average cooling efficiency of 62.6%, while the circular hole design reaches 60.4%. For the slot hole scheme, the inner-side and outer-side inflow rates are 1.1 g/s and 1.3 g/s, respectively, whereas the circular hole scheme has lower inflow rates of 0.84 g/s and 1.1 g/s. Consequently, the circular hole design reduces cooling air consumption by 19.2%.

The circular hole cooling design is integrated into the overall flow field, including the rotating detonation flow, for numerical simulation. Figure 10(a) presents the outer endwall temperature contours, revealing that high-temperature regions are primarily concentrated near the detonation wave and oblique shock wave, with temperatures gradually decreasing downstream. Before reaching the turbine blade, the oblique shock wave first interacts with the film cooling region. After mixing with the secondary flow, the gas temperature drops below 2000 K, nearly halving compared to the uncooled mainstream temperature. Significant cooling also occurs near the leading edge, pressure side, and suction side.

Figure 10(b) illustrates the pressure gradient on the outer endwall. Compared to the uncooled case in Figure 7, the pressure gradient is concentrated near the oblique shock wave and film cooling holes. Downstream and between the turbine blades, the pressure gradient remains minimal since the oblique shock wave first impinges on the film cooling holes. The porous structure and secondary flow mixing mitigate intense pressure fluctuations.

To more intuitively quantify the temperature distribution on the blade surface, both with and without film cooling holes, the average temperature along the axial direction of the RDC for the blade (from hub to tip) is calculated and presented in Figure 11. Overall, the proposed endwall cooling scheme effectively reduces thermal loads on turbine blade tips and roots while alleviating pressure gradients between blades. This not only prevents localized pressure concentrations but also extends turbine blade service life.

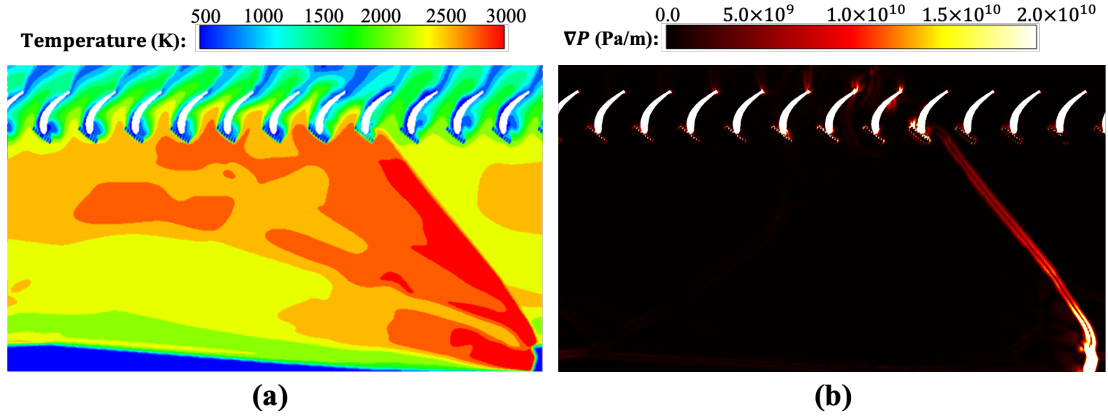

Figure 10. Outer endwall surface contour plot with film cooling holes: (a) Temperature and (b) Pressure gradient.

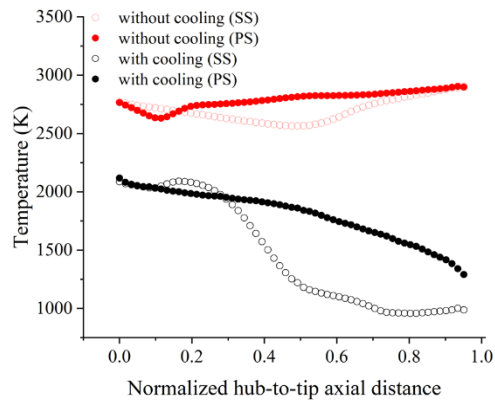

Figure 11. Average blade-surface temperature distribution along the axial direction of the RDC, plotted against the normalized hub-to-tip distance of the blade.

The local flow field on the outer wall is analyzed, and the results are shown in Figure 12. When the oblique shock wave initially sweeps over the film cooling holes, it suppresses the outflow of the secondary flow. This occurs because the oblique shock wave not only carries high temperatures but also causes a localized pressure increase, compressing the cooling gas back into the plenum. Although this suppression reduces instantaneous cooling efficiency, the cooling system still reduces the temperature of the post-shock gas, originally exceeding 3000 K, to below 2000 K before it reaches the turbine blade. As the oblique shock wave continues to propagate, the cooling effect of the secondary flow recovers.

To compare with the circular hole design, the performance of the slot hole design in the overall flow field is also tested. The inflow parameters of the circular hole and slot hole are listed in Table 2.

Table 2. Inflow parameters of the circular hole and slot hole

|  | Mass flow rate(kg/s) | Surface area (mm$^2$) | blowing ratio |
|---|---|---|---|
| Main flow | 0.1 | 175 | / |
| Secondary flow(slot) | 0.0134 | 33.456 | 0.7 |
| Secondary flow(hole) | 0.0084 | 22.276 | 0.66 |

It is found that slot holes provide excellent cooling in localized regions. However, the secondary flow creates a cold vortex on the pressure side of the turbine blade. While this vortex enhances cooling efficiency in specific areas, it also reduces the overall coverage of the secondary flow. In summary, circular holes are characterized by lower air consumption, broader cooling coverage, and ease of manufacturing. Slot holes, on the other hand, excel in localized cooling efficiency. Considering these factors, the circular hole wall cooling design is adopted for subsequent research.

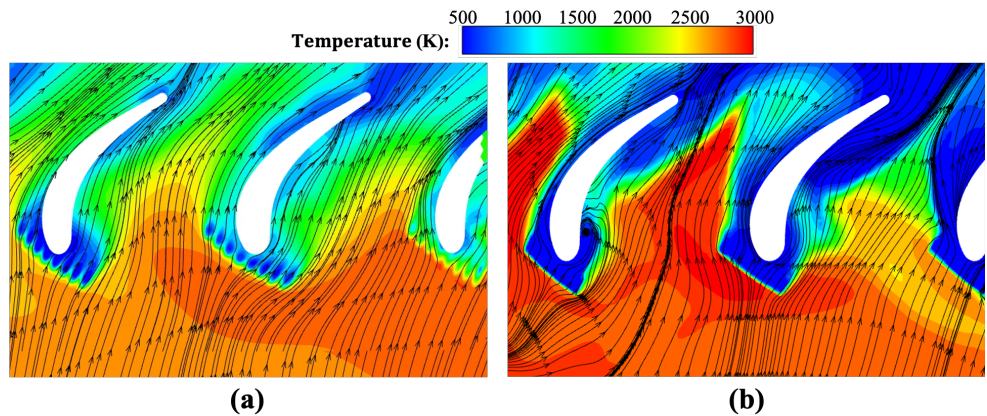

Figure 12. Temperature contour plot of endwalls with film cooling holes: (a) Slot holes and (b) Circular holes.

Since this study performs a 3D simulation of the flow field, the temperature distribution along the blade height is also analyzed, in addition to the regions near the blade tip and root. Figure 13 illustrates the temperature characteristics of the blades in the detonation flow field. The blade in contact with the oblique shock wave is designated as $B_o$, the first blade downstream of the shock wave as $B_{O+\Delta}$, the second blade as $B_{O+2\Delta}$, and so on. The blade temperatures downstream of the shock wave peak and gradually decrease. The cooling performance of the blade surfaces in the detonation flow field resembles the results from the single-blade simulation shown in Figure 9. Specifically, cooling at the blade leading edge is primarily concentrated near the root and tip. However, unlike the single-blade simulation, the cooling effect on the blade sides and trailing edge diffuses along the blade height. This is due to the pulsating nature of the flow field in the detonation combustor, which enhances thermal exchange between the secondary flow and the main flow. Consequently, high-temperature regions are mainly concentrated at the leading edge of the turbine blades downstream of the oblique shock wave.

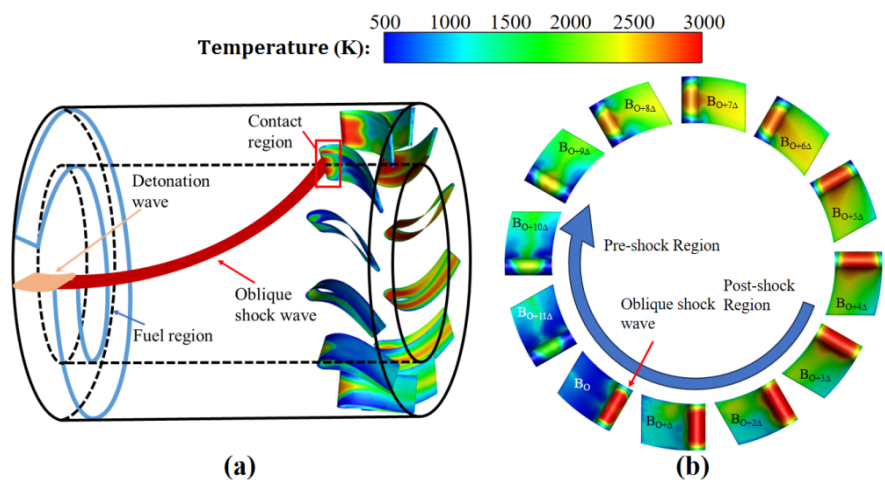

Figure 13. Temperature contour plot of the full-span blade with endwall film cooling holes: (a) XZ-plane view and (b) XY-plane view.

## 3.3 Flow field analysis with endwall and leading-edge film cooling on the turbine blades

As shown in the previous results, although endwall film cooling effectively protects the blade tip and root regions, the leading-edge area of the blades remains a high-temperature region due to the influence of the high-temperature and high-pressure downstream flow field from the detonation combustor. To address this issue, this study proposes a combined film cooling scheme for both the endwalls and the leading edge.

To validate the feasibility of the leading-edge film cooling scheme and evaluate the cooling performance under different conditions, a steady-state numerical simulation is conducted on a single blade with leading-edge film cooling. The arrangement of film cooling holes for the vertical and vertical-inclined scheme is shown in Figure 3. Three rows of holes—Pressure-Side Row 1, Suction-Side Row 1, and Suction-Side Row 2, from left to right—have a diameter of 0.4 mm. Monitoring lines are placed between adjacent rows of film cooling holes to quantify the cooling efficiency under various scenarios. The secondary flow plenum is set with mass flow rate inlet conditions of 1 g/s, 3 g/s, and 5 g/s. Figures 14 and 15 show the visualization of film cooling efficiency for both configurations under these mass flow rates.

The results show that between Pressure-Side Row 1 and Suction-Side Row 1, the vertical-inclined scheme provides significantly higher cooling efficiency and broader coverage compared to the vertical scheme. Between Suction-Side Row 1 and Suction-Side Row 2, the difference between the two configurations is relatively smaller. The average cooling efficiency over the blade surface is calculated for each condition. At a mass flow rate of 1 g/s, the Vertical Scheme achieves 77.23%, while the vertical-inclined scheme achieves 79.17%. At 3 g/s, the Vertical Scheme reaches 80.86%, and the vertical-inclined scheme reaches 82.24%. At 5 g/s, the Vertical Scheme achieves 82.25%, while the vertical-inclined scheme achieves 82.96%. These results indicate that the vertical-inclined scheme generally provides higher average cooling efficiency. While increasing the mass flow rate from 1 g/s to 3 g/s significantly improves cooling efficiency, the improvement from 3 g/s to 5 g/s is less pronounced. This is because excessive cooling airflow causes the secondary flow to detach from the wall surface, reducing the cooling efficiency. It is necessary to test specific operating conditions to determine the optimal boundary conditions. For the current test, a mass flow rate of 3 g/s is found to be the most suitable for this scenario.

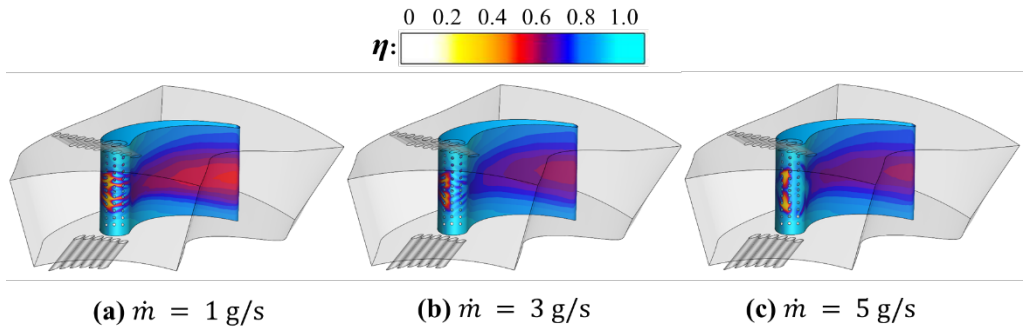

(a) $\dot{m} = 1$ g/s  (b) $\dot{m} = 3$ g/s  (c) $\dot{m} = 5$ g/s

Figure 14. Cooling efficiency under different inlet flow conditions for the vertical scheme.

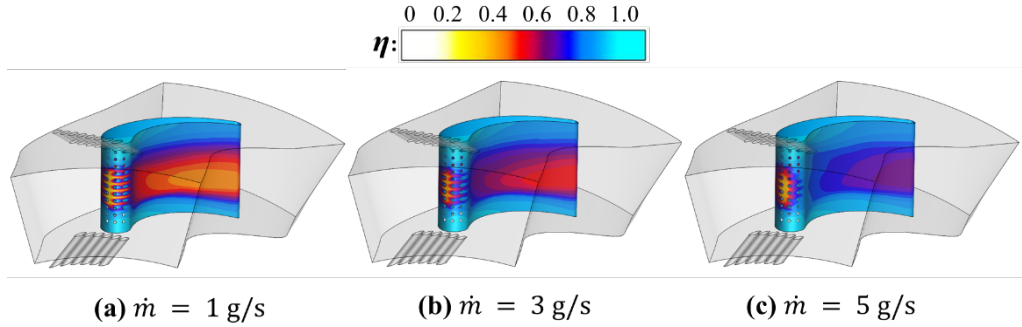

(a) $\dot{m} = 1$ g/s    (b) $\dot{m} = 3$ g/s    (c) $\dot{m} = 5$ g/s

Figure 15. Cooling efficiency under different inlet flow conditions for the vertical-inclined scheme.

Since the high-temperature region of the blade is primarily located between the two rows of film cooling holes on the leading edge, monitoring lines are placed between these rows to collect data for calculating cooling efficiency. The resulting cooling efficiency curves are shown in Figure 16. Consistent with the observations in Figure 9, the region immediately downstream of the film holes—including both the endwall and leading-edge film holes—exhibits notably high film cooling efficiency. Specifically, in the immediate vicinity downstream of the cooling hole exits, the coolant jet is freshly emitted and has yet to undergo significant thermal mixing with the mainstream hot gas. At this stage, the coolant forms an unmixed protective layer directly over the surface, maintaining a temperature close to its initial state and resulting in a cooling efficiency approaching unity. Meanwhile, in the overlapping zones where adjacent coolant jets intersect, multiple rows of film holes arranged circumferentially produce coolant jets that merge downstream. This merging increases the film thickness and enhances the continuity of coolant coverage, thereby also maintaining substantially high cooling efficiency.

From the cooling efficiency curves, it is also evident that the vertical-inclined scheme achieves significantly higher cooling efficiency between the pressure side and the suction side compared to the vertical scheme. Additionally, the cooling efficiency between the two suction sides is generally slightly higher for the vertical-inclined scheme than for the vertical scheme. Although two regions of low cooling efficiency appear under the 5 g/s condition for the vertical-inclined scheme, the areas of these low-efficiency regions are significantly smaller compared to those observed in the vertical scheme.

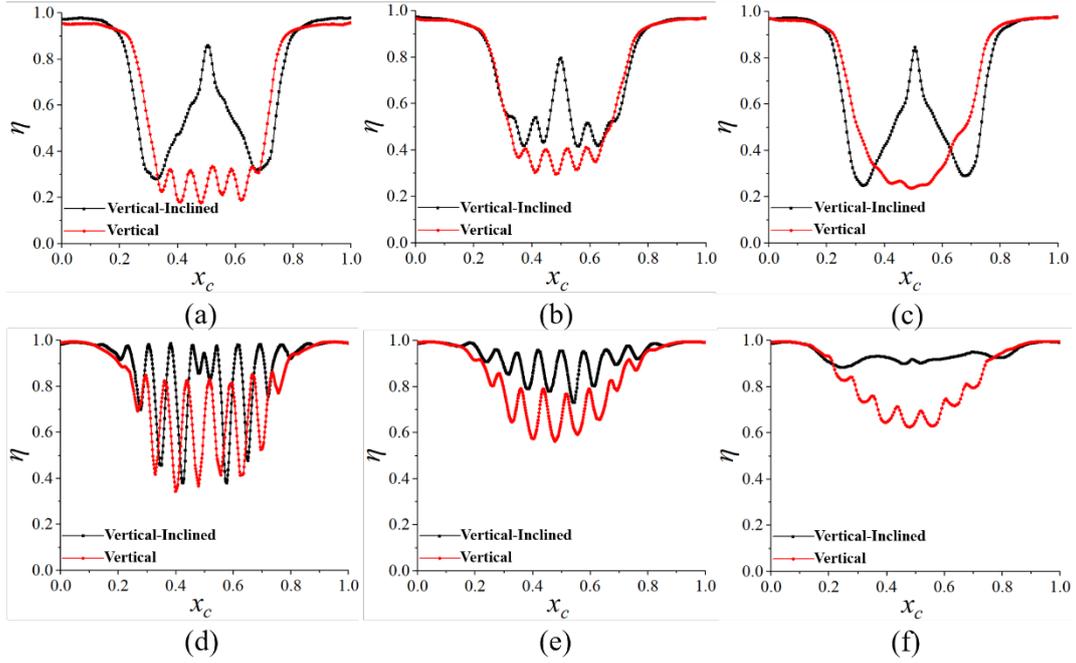

Figure 16. Cooling efficiency along data extraction lines: (a) Line 1 ($\dot{m}$=1 g/s), (b) Line 1 ($\dot{m}$=3 g/s), (c) Line 1 ($\dot{m}$=5 g/s), (d) Line 2 ($\dot{m}$=1 g/s), (e) Line 2 ($\dot{m}$=3 g/s), (f) Line 2 ($\dot{m}$=5 g/s).

In this study, a cloud plot is created by taking a cross-section through the center axis of the middle hole row (Figure 17) to observe the mixing of secondary flow and mainstream. In the vertical scheme, the secondary flow forms a cold air vortex after being injected into the mainstream, with a height of approximately 1.7d. Analyzing the data along the intersection line of this cross-section and the blade, it is found that the region where the cooling efficiency exceeds 80% accounts for 51.83% of the total area length, and the region where the cooling efficiency exceeds 95% accounts for 37.16%. The cold air vortex causes the secondary flow coverage area to rise, moving further from the wall surface, which prevents the cold air from uniformly covering the hot surface and negatively impacts the cooling efficiency. In contrast, the vertical-inclined scheme shows a clear advantage in secondary flow adhesion. After injection into the mainstream, except for the center vertical hole, the height of the cold air film in the other holes is about 1.3d. The region where the cooling efficiency exceeds 80% occupies 99.1% of the total area length, and the region where the cooling efficiency exceeds 95% occupies 64.96%. Overall, in the steady-state single-blade simulation, the vertical-inclined scheme significantly outperforms the vertical scheme in terms of cooling efficiency.

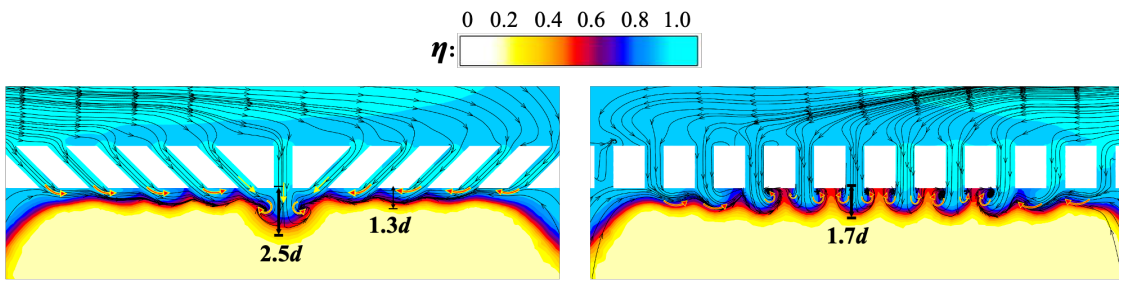

Figure 17. Streamline plot of the flow field near film cooling holes: (a) Vertical and (b) Vertical-inclined.

Based on the single-blade simulation results, a preliminary evaluation of the two film cooling hole arrangements can be made. Although the inlet boundary conditions in the above cases are adjusted to approximate the average or extreme values of the detonation flow field outlet, the simulations essentially represent steady-state flow fields, ignoring the time-dependent effects in the flow field—such as those caused by shock waves. This condition is more akin to a traditional combustion chamber and aims to significantly reduce computational costs while providing an initial test of the model. Additionally, it serves as a reference for comparing results obtained from the coupled detonation flow field, thereby revealing the impact of detonation waves on film cooling performance.

To further investigate the performance of film cooling in unsteady flow fields, simulations are conducted on a coupled flow field that includes the turbine region and the rotating detonation combustion chamber. The endwall film cooling scheme is employed, with circular film cooling holes on the inner and outer endwalls of the turbine region. To reduce computational costs and simplify comparisons, the stator blade row consists of 10 blades without leading-edge film cooling and 2 blades equipped with leading-edge film cooling. Two leading-edge film cooling configurations are considered: the vertical scheme and the vertical-inclined scheme. These configurations are used to evaluate the cooling performance of different leading-edge film cooling structures under detonation flow conditions.

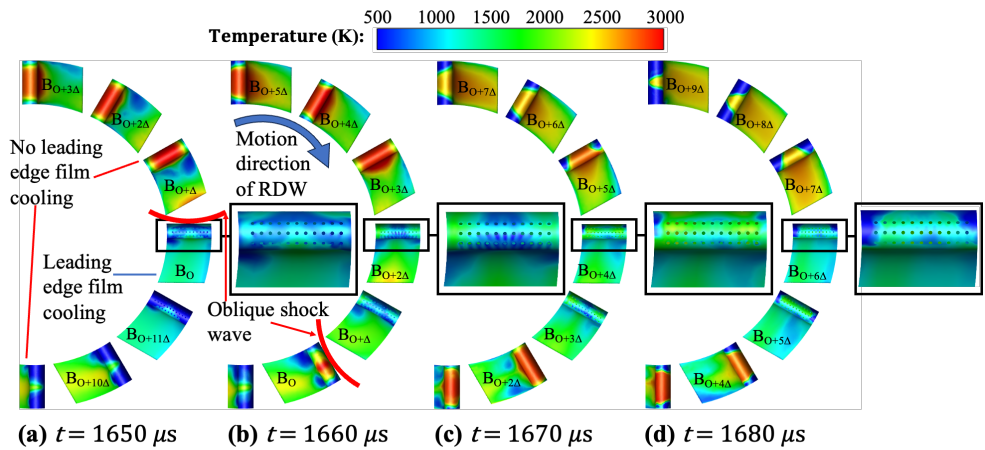

Figure 18. Temperature contour of the downstream blade flow field with vertical film cooling.

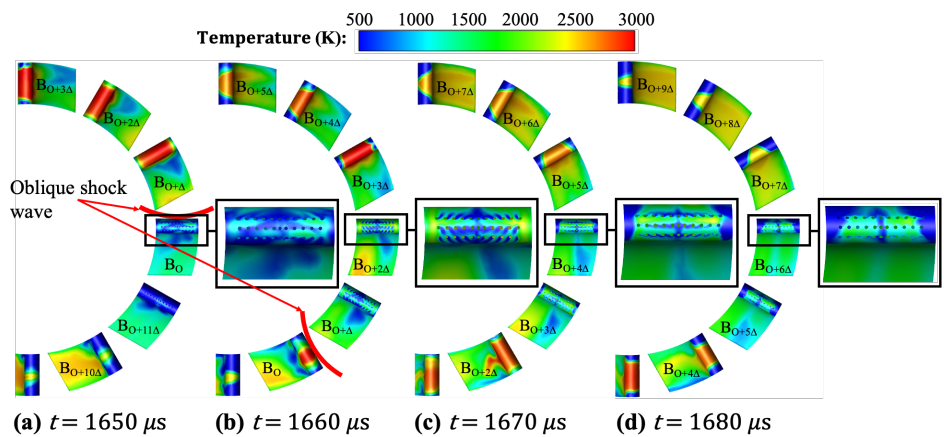

Figure 19. Temperature contour of the downstream blade flow field with vertical-inclined film cooling.

Figures 18 and 19 display the temperature distribution of turbine blades at different moments after adopting the vertical scheme and the vertical-inclined scheme at the leading edge. As shown in Figure 17(a), two blades have both endwall film cooling and leading-edge film cooling, while the remaining blades only have endwall film cooling. Previous results indicate that the detonation wave completes one rotation in about 60 μs, and 10 μs corresponds to approximately 1/6 of the cycle, which is equivalent to the movement of two blade angles. Based on the numbering method in Figure 13, the blades at different moments are numbered as shown in Figures 18 and 19. The flow field at each subsequent frame is considered the result of the previous frame after rotating 2 blade angles in a clockwise direction. For example, as shown in Figure 19(a), at 1650 μs, the temperature in Area 3 exceeds 3000 K. After 10 μs, Area 3 sweeps over two blades and moves to the endwall film cooling region shown in Figure 19(b), where the temperature of the leading edge drops below 2000 K. After another 10 μs, Area 3 moves to the position shown in Figure 19(c). After leaving the blade with leading-edge film cooling, the temperature in this region rises to around 2800 K.

By analyzing the average temperature along monitoring line 1 at the leading edge of each region and grouping them according to the numbering, the results shown in Figure 20 are obtained, where $l_b$ represents the number of Δ (i.e., 1Δ, 2Δ, etc.) from the blade to the detonation wave. As shown in Figure 20(a), the average film cooling efficiency for the vertical scheme is 45.66%, with lower cooling efficiency in the wave-back regions, especially in the 5Δ and 6Δ regions. This can be attributed to the accumulation of cold air at the cooling hole exit when the film cooling zone is located ahead of the shock wave. As the shock wave moves, the accumulated cold air is carried away, resulting in poor cooling efficiency when the wave-back region passes through the leading-edge film cooling area. In contrast, the vertical-inclined scheme (Figure 20 (b)) shows an average film cooling efficiency of 50.32%. The amount of cold air accumulated in the film cooling region before the wavefront is smaller, but the cold air adheres better, making it less influenced by the movement of the shock wave and providing better overall cooling for the blade leading edge. Furthermore, compared to the results from the steady flow field, it can be observed that the detonation flow field is more favorable for the mixing and diffusion of secondary flow in the mainstream to some extent.

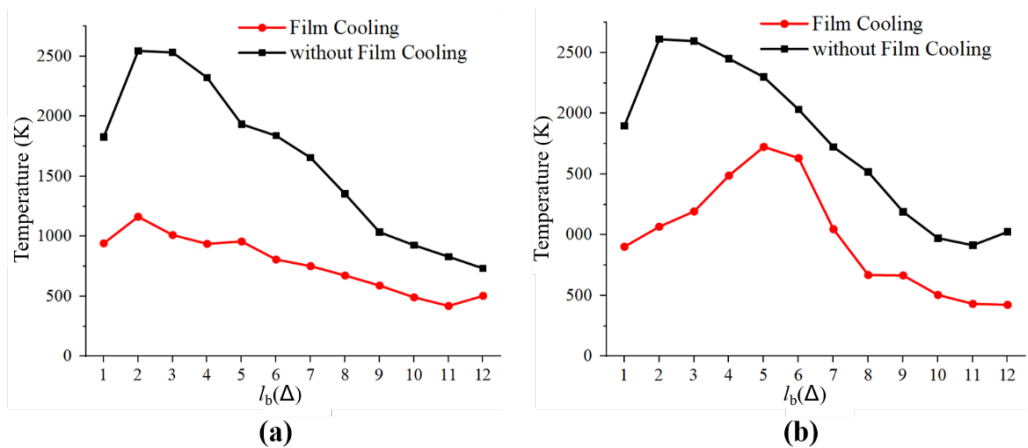

Figure 20. Temperature distributions of the flow fields: (a) Vertical scheme and (b) Vertical-inclined scheme.

## 4. Conclusion

This paper presents a numerical simulation study of turbine blade film cooling in an RDC-turbine coupled flow field. Slot and circular holes are employed for endwall cooling, while vertical and inclined hole configurations are applied for leading-edge film cooling. The results demonstrate that the combination of endwall cooling and leading-edge film cooling effectively reduces the blade surface temperature. The main conclusions are as follows:

1. Endwall film cooling on the turbine blades alleviates the thermal load on the endwalls, reducing the peak temperature of the post-detonation products from over 3000 K to below 2000 K before they reach the turbine blades. It also reduces the formation of shock waves between blades and enhances flow stability. In cases with similar performance, the circular hole reduces cooling air consumption by 19.2% compared to the slot hole, confirming the applicability of circular holes for turbine endwall cooling.
2. Leading-edge film cooling effectively reduces high temperatures at the blade leading edge. The vertical-inclined scheme has higher cooling efficiency than the vertical scheme, with improved secondary flow attachment. Additionally, in the detonation flow field, where periodic and pulsating effects influence the mainstream, the vertical-inclined scheme exhibits greater stability. The combination of endwall cooling and leading-edge film cooling provides comprehensive thermal protection for turbine blades in such a flow environment.
3. Furthermore, an examination of the flow fields in film-cooled turbine blades, both with and without the influence of a rotating detonation wave, demonstrates that the detonation wave intensifies secondary flow mixing and mainstream diffusion. This interaction significantly impacts cooling effectiveness, emphasizing the importance of optimizing film cooling strategies for RDC-based turbine systems.

**APPENDIX A: Validation of the one-step hydrogen mechanism with detailed chemistry**

While it is true that a more detailed chemical mechanism could provide a more accurate prediction of, for example, the flame structure, intermediate species, the overall flow-field simulations using the one-step chemistry model are in very close agreement with experimental and theoretical expectations, particularly for supersonic combustion where the flow time scale is typically on the order of $\sim 10^{-6}$ s [61]. This indicates that, despite the simplifications in the reaction mechanisms, the model captures the dominant physical processes governing the RDC flow field, which is needed for the current study of a complex three-dimensional RDC–turbine system, and the current study does not aim to resolve the fine cellular structure of the detonation wave.

That said, we evaluate the applicability of the one-step hydrogen mechanism with a detailed chemistry model (9 species and 19 steps) by Ó Conaire et al. [62]. Simulations are conducted using a two-dimensional detonation tube, with each reaction mechanism adopted for the case. The detonation tube has geometric dimensions of 400 mm in length and 60 mm in width. The boundary conditions are as follows: the right side is set as a pressure outlet at 1 atm, while the remaining walls are treated as adiabatic boundaries. The flow field is initialized with a temperature of 300 K and a pressure of 0.1 MPa. Three ignition zones, with a pressure of 2 MPa and a temperature of 2000 K, are located near the left wall, and the hydrogen equivalence ratio in the remaining regions is set to 1. The pressure contours and flow field profiles (pressure and temperature) are shown in Figure 21, and Figure 22, respectively. The overall characteristics of the detonation waves are consistent between the one-step and detailed chemistry solutions. Although small discrepancies occur near the peak values, the deviations remain within a reasonable range. The differences in propagation velocities between the two approaches are also small, with deviations of approximately 5% for the one-step solution and 1% for the detailed chemistry solution. Given the high computational cost of using detailed chemistry in a three-dimensional simulation of the RDC, the one-step chemistry model is adopted in the present study. Moreover, as further confirmed in Appendix B, the one-step chemistry model accurately predicts the thrust performance of the RDC, showing close agreement with the experimental data.

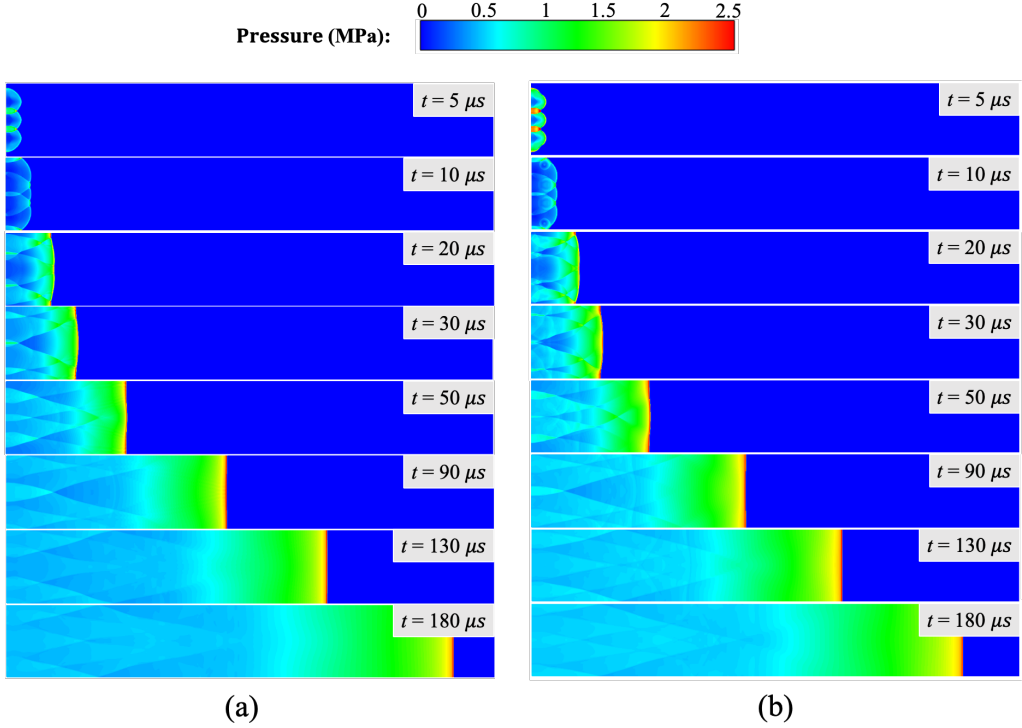

Figure 21. Pressure contour plots of the detonation tube solutions at different snapshots: (a) One-step chemistry and (b) Detailed chemistry [62].

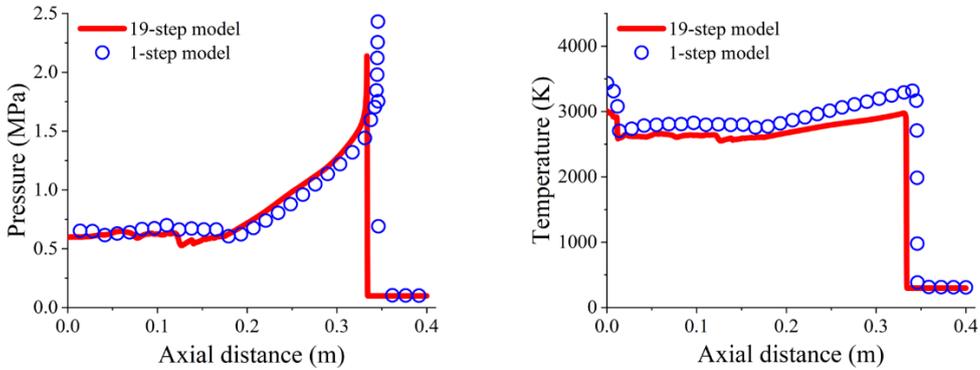

Figure 22. Pressure and temperature profiles of the detonation tube solutions at different snapshots.

**APPENDIX B: Validation of the RDC simulation with experimental thrust data**

To further confirm the reliability of the simulation setup proposed in this paper, validation is performed for the RDC by comparing the results with experimental data (i.e., specific impulse) from various previous studies [63, 64], as well as with the analytical solution by Wintenberger et al. [65]. Nine operating conditions, corresponding to equivalence ratios of 0.6, 0.8, 1.0, 1.2, 1.4, 1.6, 1.8, 2.0, and 2.2, are considered. The fuel-based specific impulse for each condition is calculated following Fotia et al. [64] by $I_{sp} = \frac{F_{thrust}}{\dot{m}_{fuel} \cdot g_0}$, where $F_{thrust}$ is the total thrust, $\dot{m}_{fuel}$ is the mass flow rate of fuel (hydrogen), and $g_0$ is the gravitational acceleration. The simulation results are presented in

Figure 23. As shown in the figure, the specific impulses obtained from the present simulations are in good agreement with both the experimental data and the analytical solution, thereby further validating the numerical methods employed in this study.

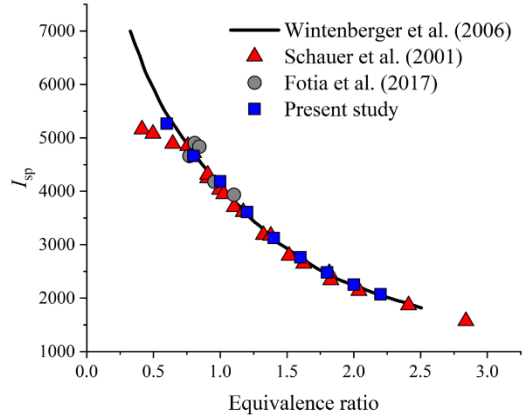

Figure 23. Comparison of the specific impulse for the hydrogen-air RDC under different equivalence ratios.

**APPENDIX C: Validation of turbulence and film cooling models with benchmark experimental data**

To verify the validity of the film cooling settings and the reliability of the simulation results, a validation study is performed using a flat-plate film cooling model. The experimental setup of Li et al.[66] serves as the benchmark. In the simulation, three blowing ratios ($M$) of 0.5, 1, and 2 are selected, covering the range considered in the present study. For each blowing ratio, both the k–ε and k–ω turbulence models are used for calculation. The data collection area is nondimensionalized using the hole diameter $d$ of 2 mm, following the method described in Ref. [66]. The spacing between holes in the same row is $10d$, and the spacing between holes in the same column is $6d$, with the airflow direction defined as the positive $x$-axis. Based on these parameters, the average film cooling efficiency curve along the $x$-direction is plotted. The comparison results (Figure 24) indicate that, except at $M=0.5$, where the $k$–$\varepsilon$ and $k$–$\omega$ models produce similar results, the $k$–$\varepsilon$ model shows better agreement with the experimental data for the other blowing ratios. This confirms the reliability of using the k–ε model in this study. Furthermore, the temperature contour plots obtained from the simulation closely match the experimental measurements, further verifying the validity of the overall case settings for film cooling.

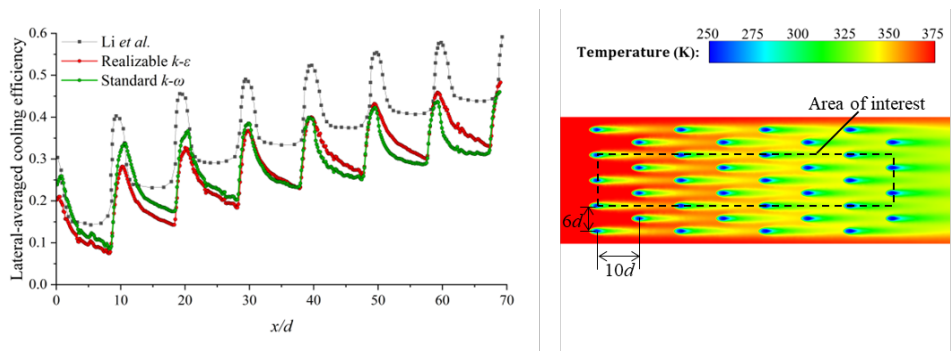

**(a)** $M = 0.5$

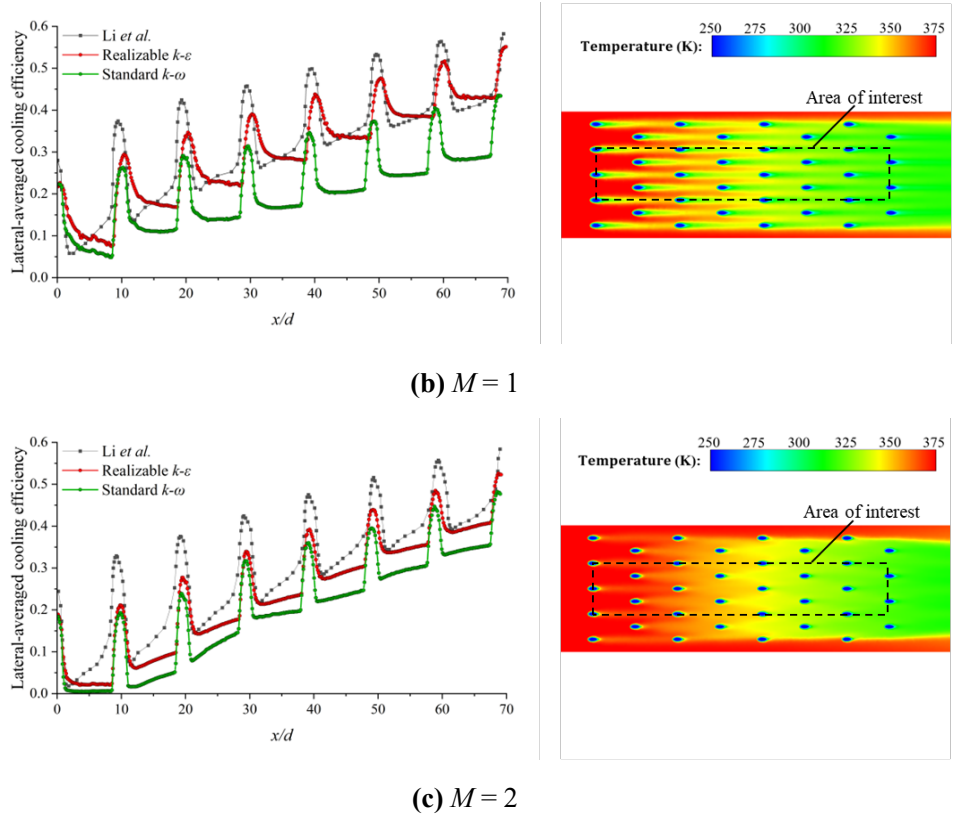

**(b)** *M* = 1

**(c)** *M* = 2

Figure 24. Average film cooling curves and temperature contour plots of the flat plate under different blowing ratios.